\documentstyle[11pt,newpasp,epsf]{article}

\begin{document}

\title{Stellar abundance analyses in the light of 3D 
hydrodynamical model atmospheres}

\author{Martin Asplund}
\affil{Research School of Astronomy and Astrophysics, 
Mt Stromlo Observatory, Cotter Road, Weston, ACT 2611, Australia}

\begin{abstract}

I describe recent progress in terms of 3D hydrodynamical model
atmospheres and 3D line formation and their
applications to stellar abundance analyses of late-type stars.
Such 3D studies remove the free parameters inherent in classical
1D investigations (mixing length parameters, macro- and microturbulence)
yet are highly successful in reproducing a large arsenal of
observational constraints such as detailed line shapes and asymmetries.
Their potential for abundance analyses is illustrated by discussing
the derived oxygen abundances in the Sun and in metal-poor stars,
where they seem to resolve long-standing problems as well as
significantly alter the inferred conclusions.

\end{abstract}


\section{Introduction}

Determining stellar element abundances play an integral role in 
most endeavours to improve our understanding of stellar, galactic
and cosmic evolution.
The term {\em observed abundances} is somewhat of
a misnomer however, since the chemical composition of course can not 
be inferred directly from an observed spectrum.
The obtained abundances are therefore never more 
trustworthy than the models employed
to analyse the observations.
The derivation of accurate abundances require realistic models
of both the spectrum formation region and the spectrum
formation process, which is the topic of the present contribution.

For a long time, understanding stellar convection has
been one of the greatest challenges in stellar astrophysics.
Convection can provide an efficient means of transporting energy
and thereby greatly influence the stellar structure and
evolution, while also mixing nuclear-processed material
and preventing radiative diffusion and gravitational settling.
For late-type stars, the surface convection zone reaches
the stellar atmosphere, which directly affects the 
emergent spectrum. 
The solar granulation is the observational
manifestation of convection: concentrated, rapid,
cold downdrafts in the midst of broad, slow, warm upflows. 
Qualitatively similar granulation properties are expected
in other solar-type stars, as indeed confirmed by
3D numerical simulations (e.g. Nordlund \& Dravins 1990;
Asplund et al. 1999; Asplund \& Garc\'{\i}a P{\'e}rez 2001;
Allende Prieto et al. 2002)
and indicated by observed line asymmetries.
The up- and downflows have radically different
temperature structures (Stein \& Nordlund 1998),
which can not be approximated by normal theoretical
1D hydrostatic model atmospheres with different
effective temperatures $T_{\rm eff}$ (Fig. 1).
In fact, neither the steep temperature structures of
the ascending material nor the shallow gradients of
the descending gas are well described by such 
1D models. Because of the photospheric inhomogeneities
and the highly non-linear and non-local nature
of spectrum formation, it is clear that no single 
1D model can be expected to properly describe all aspects of
what is inherently a 3D phenomenon.

Here I will describe recent
progress in developing 3D hydrodynamical
model atmospheres of late-type stars and some applications to 
stellar abundance analyses. In spite of their
still inferior treatment of radiative transfer and
line blanketing compared with existing 1D model
atmospheres based on the mixing length theory, 
detailed comparisons with a range
of observational diagnostics suggest that
these simulations are indeed highly realistic.
%It seems like the many advantages with this approach
%more than compensates for the few drawbacks.
%The new generation of 3D model atmospheres 
%may thus represent an important step towards
%the goal of improving the accuracy of derived
%stellar abundances by making obsolete the traditional
%free parameters (mixing length parameters,
%micro- and macroturbulence) of 1D analyses
%and predicting detailed spectral line shapes and asymmetries,
%which agree almost perfectly with observations
%(e.g. Asplund et al. 2000b,c; Allende Prieto et al. 2002;
%Nissen et al. 2000). 

\section{3D Hydrodynamical Model Atmospheres}

The 3D model atmospheres which form the basis of
the abundance analyses presented here have been
computed with a 3D, time-dependent, compressible, explicit,
radiative-hydrodynamics code developed to
study solar and stellar surface convection
(Stein \& Nordlund 1998). The hydrodynamical
equations for conservation of mass, momentum
and energy are solved on a non-staggered Eulerian
mesh with gridsizes of $\approx 100^3$. The physical
dimensions of the grids are sufficiently large
to cover many ($>10$) granules simultaneously 
in the horizontal direction and about 13 pressure
scale-heights in the vertical. In terms of
continuum optical depth the simulations extend
at least up to log\,$\tau_{\rm Ross} \approx -5$ which for
most purposes are sufficient to avoid 
numerical artifacts of the open upper boundary on
spectral line formation. 
The lower boundary is located at large depths to
ensure that the inflowing gas is isentropic and featureless, 
while periodic horizontal boundary conditions are employed. 
The temporal evolution of the simulations cover
several convective turn-over time-scales 
to allow thermal relaxation to
be established and to obtain statistically significant 
average atmospheric structures and spectral line profiles. 
No magnetic fields have been included in the simulations
described here although the code is capable of
simulating magneto-convection.

\begin{figure}
%\plotone{asplundm_f1.eps}
\plotfiddle{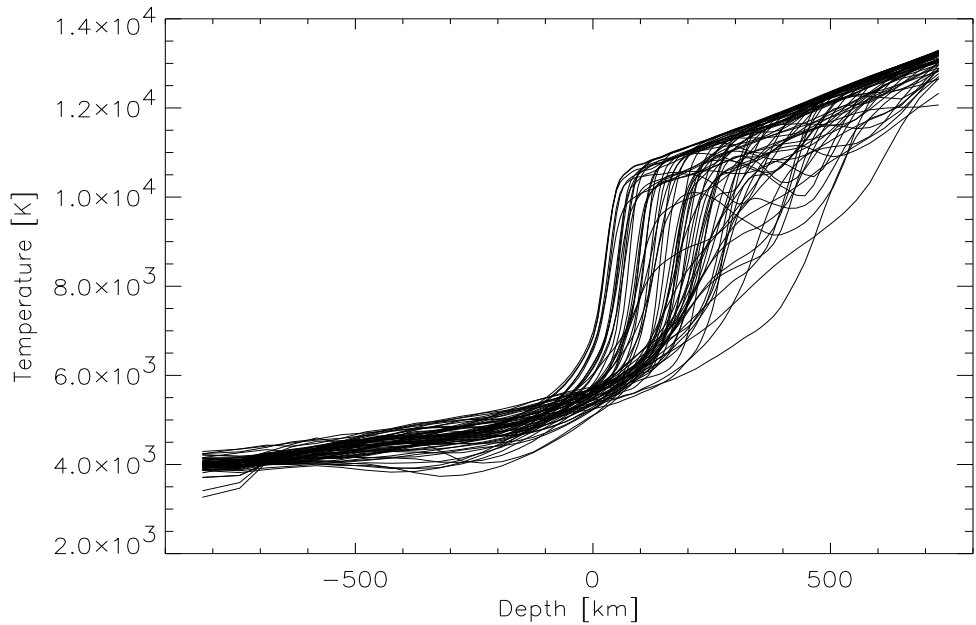}{8cm}{0}{75}{70}{-190}{0}
\caption{Temperature structure of a 3D solar model atmosphere illustrating
the vastly different temperature gradients in the up- and downflows.
Vertical columns are connected with solid lines.}
\end{figure}

In order to obtain a realistic atmospheric structure,
it is crucial to have the best possible input physics, and
properly account for the energy exchange between
the radiation field and the gas. 
The adopted equation-of-state is that of Mihalas et al. (1988),
which includes the effects of ionization, excitation
and dissociation. The continuous opacities come from
the Uppsala package (Gustafsson et al. 1975 and subsequent
updates) while the line opacities are from Kurucz (1998, private communication).
The 3D radiative transfer is solved at each time-step
using 9 inclined rays (2 $\mu$-angles and 4 $\varphi$-angles,
plus the vertical) under the assumptions of local
thermodynamic equilibrium (LTE, $S_\lambda = B_\lambda$)
and opacity binning (Nordlund 1982). 
The four opacity bins are designed to correspond to 
continuum and weak, intermediate and strong lines.
The assignment of the original 2748 wavelength points into
the different opacity bins follows from detailed monochromatic radiative
transfer calculations of the 1D averaged atmospheric structure.
The opacity binning thus includes the effects of line-blanketing
in a manner reminiscent of opacity distribution functions.
In order to improve the numerical accuracy, the radiative
transfer is solved on a finer grid and the results subsequently
interpolated back to the original grid for the hydrodynamical variables
at each timestep.
%Throughout the simulations, the accuracy of the opacity
%binning technique is verified by solving the detailed
%monochromatic radiative transfer for a 2D vertical slice
%of the simulation and typically found to agree within 1\% in
%emergent flux.

It is important to realise that {\em the simulations contain
no free parameters which are tuned to improve the agreement
with observations}. The adoption of the numerical and physical
dimensions of the simulation box is determined by practical
computional time considerations, the need to resolve
the most important spatial scales and the wish to place the artificial
boundaries as far as possible from the region of interest.
The code is stabilized using a hyper-viscosity diffusion
algorithm with parameters specified from standard
hydrodynamical test cases. These parameters are not
changed in the actual stellar convection simulation and
are therefore not freely adjustable parameters. 
It has been verified that the resulting atmospheric
structures are insensitive to the adopted effective viscosity
at the current highest affordable numerical resolution
(Asplund et al. 2000a). The input parameters 
discriminating different models are the surface gravity log\,$g$,
metallicity [Fe/H] and the entropy of the inflowing material
at the bottom boundary. The effective temperature of
the simulation is therefore a property which depends on the
entropy structure and evolves
with time around its mean value
following changes in the granulation pattern.

Further details of the 3D hydrodynamical model atmospheres
are available in Stein \& Nordlund (1998), Asplund et al. (1999, 2000a,b)
and Asplund \& Garc\'{\i}a P{\'e}rez (2001).

\section{Spectral Line Formation}

\subsection{General Description}

The 3D hydrodynamical model atmospheres described in the previous
section form the basis for the 3D spectral line formation
calculations. The Doppler shifts introduced by the convective
and oscillatory motions cause line asymmetries and shifts as well
as broaden the lines significantly. Spatially resolved 
profiles thus come in an amazing assortment as a result
of the atmospheric inhomogeneities and velocities together with
their correlations (Fig. 2).
The concepts of micro- and macroturbulence, which are
introduced in 1D analyses in order to account for the missing
line broadening, are not necessary in 3D calculations, 
as evident from the excellent agreement between
predicted and observed line profiles (Fig. 3, see also
Asplund et al. 2000b) and the lack of significant trends
in derived abundances with line strength (Asplund et al. 2000c;
Asplund 2000). 
The main component of microturbulence does not arise
from microscopic turbulent motions but from gradients in
convective motions that are resolved with the present simulations,
which is also the main physical explanation for macroturbulence.
%The line formation process is heavily biased towards upflows
%(large area coverage, high continuum intensity, steep temperature
%gradients) where the divergent nature of the flow prevents
%large amount of turbulence to develop in spite of the
%high Reynolds-numbers of the plasma (Nordlund et al. 1997).

\begin{figure}
\plotfiddle{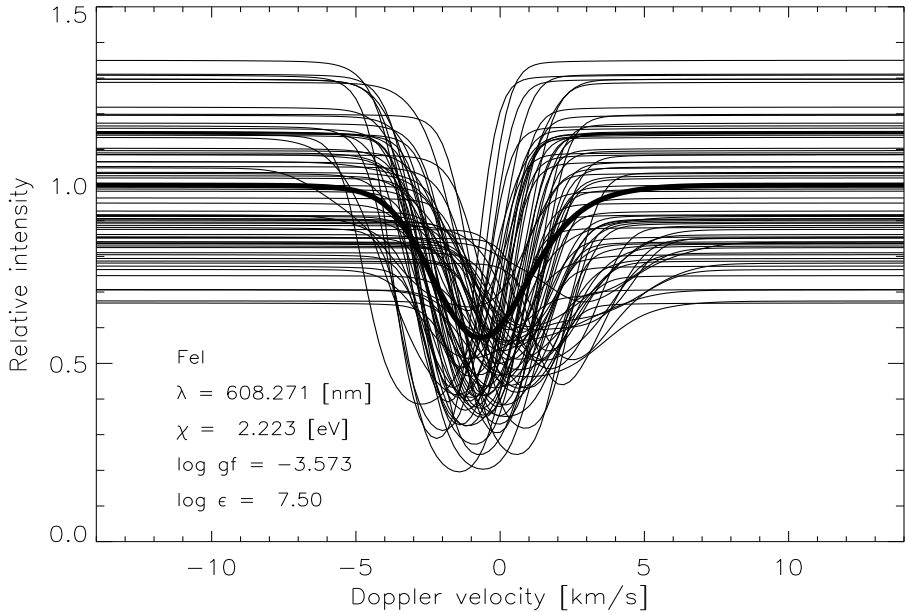}{8cm}{0}{75}{70}{-190}{0}
\caption{Predicted spatially resolved profiles of the Fe\,{\sc i} 608.2\,nm line
in the solar simulation. Note the bias of stronger lines in the upflows
due to their steeper temperature gradients in general.
%The resulting spatially and temporally averaged profile is clearly an extremely
%sensitive probe of the statistical properties of the photospheric temperatures
%and velocities.
}
\end{figure}

The original 3D hydrodynamical simulations which includes
deep, extremely optically thick layers are interpolated to
a finer depth scale which extends only down to 
log\,$\tau_{\rm Ross} \approx 2.5$
prior to the spectral line calculations for improved numerical accuracy. 
At the same time, the simulations are interpolated to a coarser
horizontal grid to ease
the computational burden. Extensive tests have ensured that
this practice is acceptable even when considering fine details
such as line asymmetries; for abundance analysis purposes the
procedure introduces non-noticable differences ($< 0.01$\,dex
for individual snapshots and much less for temporal averages).
Depending on whether the 3D line formation is computed in
LTE or in non-LTE, different assumptions and approximations 
are necessary, which are described next.

\subsection{3D LTE Line Formation}

With the assumption of LTE ($S_\lambda = B_\lambda$), 
the level population is uniquely determined by the local
gas temperature from the Boltzmann and Saha distributions.
With the source function known, it is then straightforward
to solve the radiative transfer equation, which  
typically is performed for about 100 wavelength
points across the line profile for 10-20 inclined outgoing rays for
more than 50 different snapshots. 
%A long temporal sequence is necessary
%since the self-induced oscillations
%in the simulations (which corresponds to stellar $p$-modes)
%cause regularly varying line-shifts and additional
%broadening of lines (macroturbulence in 1D terminology,
%i.e. changes the shapes but not the strengths of lines).
This allows asymmetries and shifts to be determined to
an accuracy on the m\,s$^{-1}$-level.
For abundance determinations even individual snapshots give
results accurate to within about 0.03\,dex, since at any
instant many granules are covered by the simulation box.
All in all, a single temporally and spatially averaged flux profile in 3D
correspond most of the time to
$N_{\rm t}*N_{\rm x}*N_{\rm y}*N_{\rm angles}*N_\lambda \ga 10^8$
1D radiative transfer calculations.
In addition, each 3D profile is normally computed for 
at least three different abundances to enable interpolation
to the requested line strength.
Even then, such 3D LTE line calculations are achievable
on current workstations thanks to efficient numerical
algorithms and implementations.

\begin{figure}
%\plottwo{asplundm_f3a.eps}{asplundm_f3b.eps}
\plotfiddle{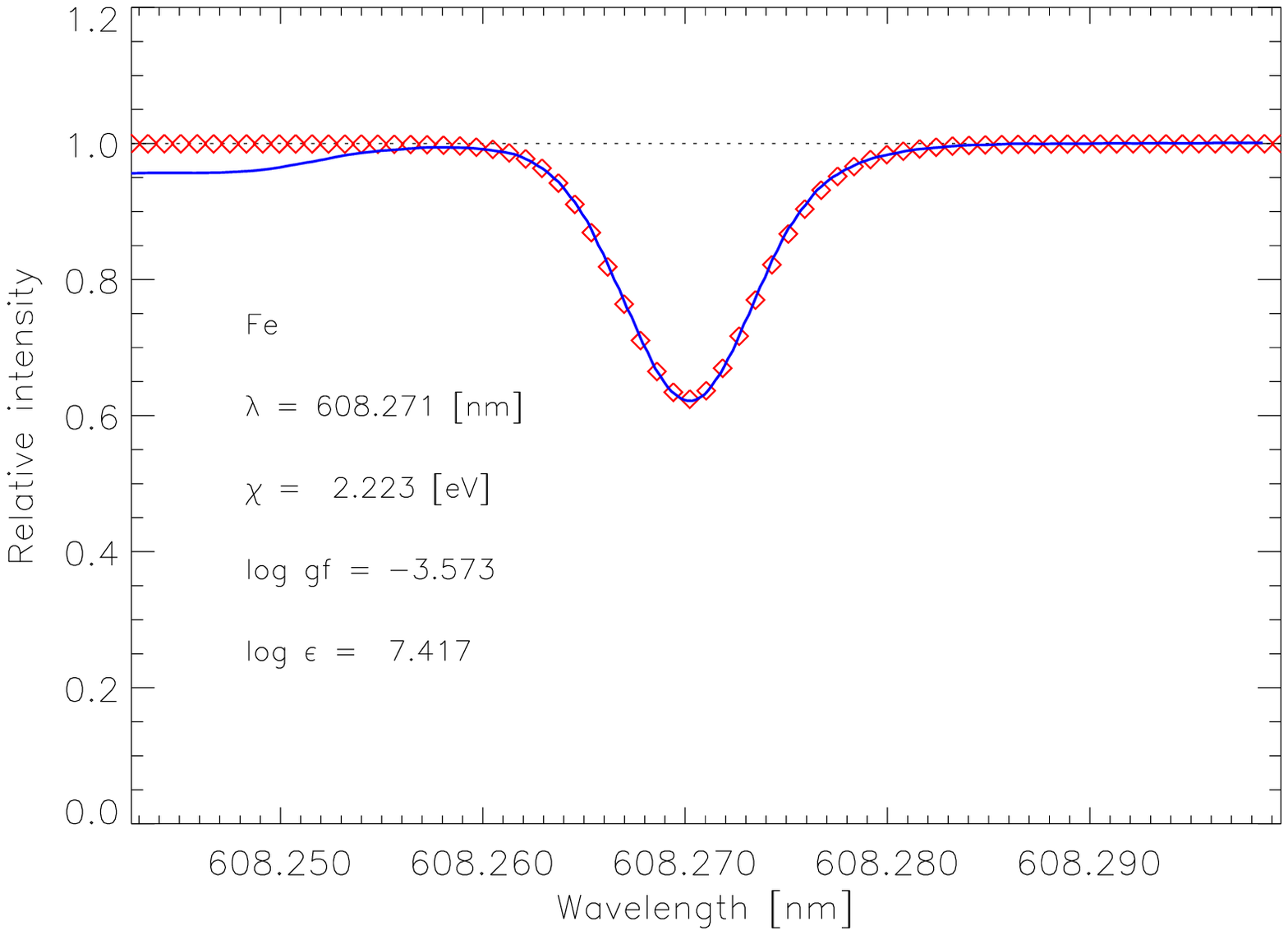}{8cm}{0}{75}{70}{-190}{0}
\plotfiddle{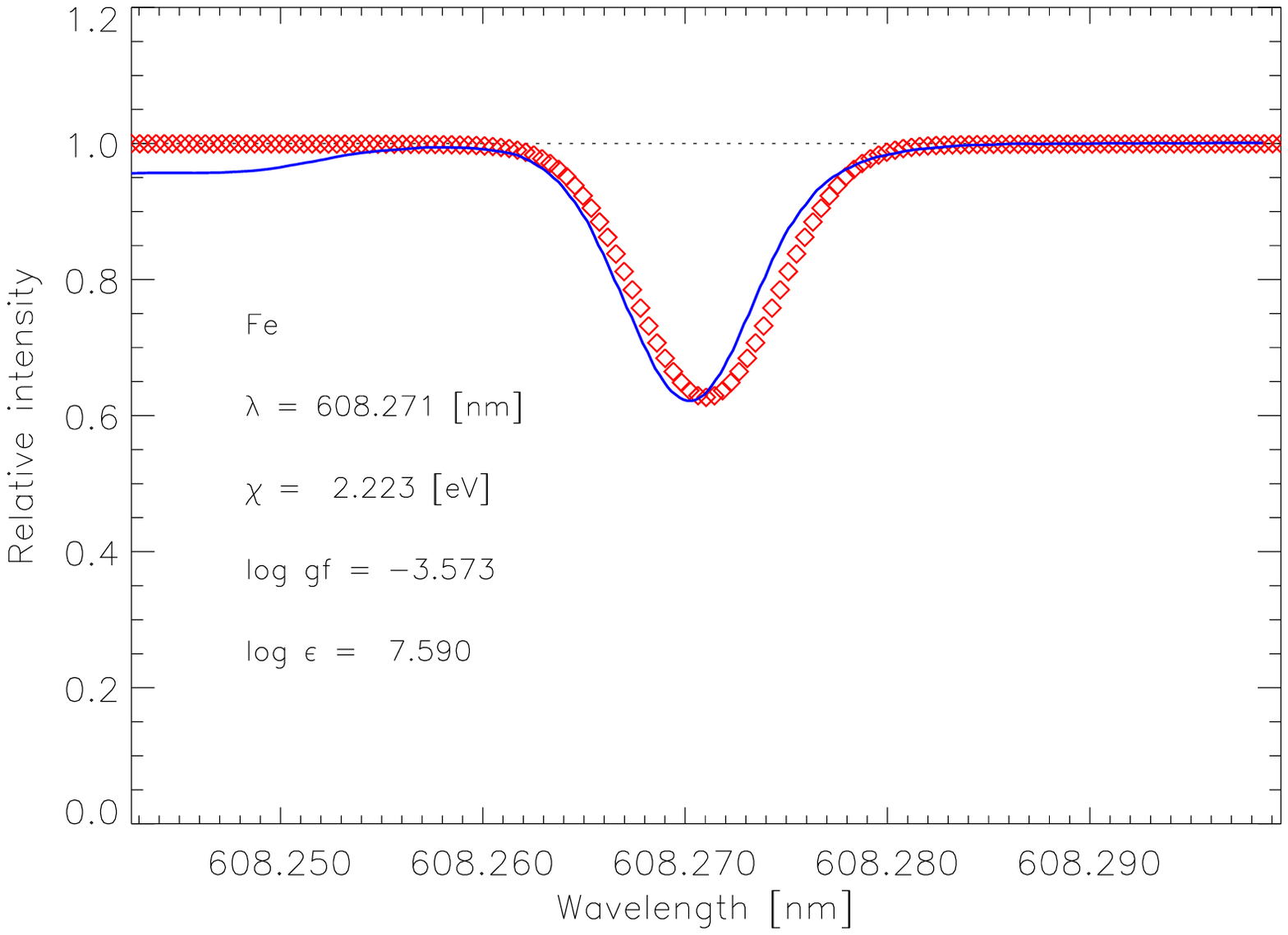}{8cm}{0}{75}{70}{-190}{0}
%\plotone{asplundm_f3.eps}
\caption{{\em Upper panel:} Comparing the predicted temporally and spatially
averaged intensity profile of the Fe\,{\sc i} 608.2\,nm
line (diamonds) with the observed solar profile (solid line).
{\em Lower panel:} Corresponding 1D profile using the
Holweger-M\"ullermodel atmosphere
with an optimized macroturbulence
(Gaussian since radial-tangential macroturbulence is not
applicable for intensity profiles) and a microturbulence of 1.0\,km\,s$^{-1}$.
Note the inability of the 1D case to predict the line shift and asymmetry,
as well as the different required Fe abundance ($+0.17$\,dex)
for the same line strength.
}
\end{figure}

No micro- and macroturbulence enter the calculations but
for flux profiles the projected stellar rotational velocity
$v_{\rm rot} {\rm sin} i$ causes line broadening in addition to the
natural, pressure and convective broadening.
Flux profiles are obtained after the actual radiative transfer
calculations from a disk-integration of the stored profiles of the
individual inclined rays for a given $v_{\rm rot} {\rm sin} i$
(Dravins \& Nordlund 1990).
The detailed comparison with observations can then proceed
through a $\chi^2$-analysis by interpolating the computed
profiles with varying abundances and adopting different
$v_{\rm rot} {\rm sin} i$.

\subsection{3D Non-LTE Line Formation}

Recently, even detailed 3D non-LTE line formation calculations
have become feasible
%with reasonably extended model atoms
(Kiselman 1997;
Uitenbroek 1998; Botnen \& Carlsson 1999; Asplund et al. 2002).
Without the much simplifying assumptions of Saha and
Boltzmann distribution for the level population, the
rate equations assuming statistical equilibrium (${\rm d}n_{\rm i}/{\rm d}t=0$)
must be solved simultaneously with the radiative transfer equation.
Due to the much increased computational effort in achieving this,
sacrifies must be made in other areas, most notably in terms of
number of snapshots the calculations can be made for.
With individual snapshots it is still feasible to carry
out calculations for atoms consisting of hundreds of radiative transitions,
each with 50-100 wavelength points.
Fortunately, with 3D model atmospheres the non-LTE effects
are expected to be quite similar between different snapshots due
the relatively large area coverage always containing many
up- and downflows, which is also confirmed by detailed calculations.
The by far dominating part of 3D non-LTE computations of
reasonably sized model atoms is the actual radiative transfer
solution from the current estimate of the source function.
This formal solution of the radiative transfer equation is performed
for all wavelengths of all transitions for typically 24-48
inclined rays to compute mean intensities and flux profiles.
The final converged solution is then obtained by a
iterative procedure, normally employing an accelerated
lambda iteration technique or similar, allowing convergence
typically within 10 iterations.
For small model atoms, the calculations are possible
to perform on modern workstations but for more extended atoms the
use of supercomputers are better suited, not the least due
to vast memory requirements with 3D non-LTE codes such
as {\sc multi3d} (Botnen \& Carlsson 1999; Asplund et al. 2002).

%      Nx Ny Nt Nmu Nv Nabu Ntran Nit
%LTE:  50x50x50x20x100   x3    x1   x1=7x10^8
%NLTE: 26x26x1 x24x60    x1    x100x10=9x10^8

\section{Stellar Abundance Analyses with 3D Model Atmospheres}

The realistic 3D hydrodynamical model atmospheres of late-type stars
together with corresponding LTE and non-LTE line formation tools described above
have a natural application to stellar abundance analyses.
For abundance studies, the different mean photospheric structure
between 3D and 1D models, the existence of atmospheric inhomogenities
and the broadening of spectral lines due to convective motions are
all important ingredients for the line formation.
To illustrate the possible impact of the new generation of 3D model
atmospheres on abundance analyses, I will describe the situation for
the most abundant cosmic metal, oxygen, in two important types of stars:
the Sun and metal-poor halo stars.
In both cases it seems like the adoption of 3D model atmospheres drastically
affects the conclusions.

\subsection{Oxygen in the Sun}

The solar oxygen abundance ought to be well determined but astonishingly
the commonly adopted value may be in error by almost a factor of two.
For the Sun, five different types of diagnostics are potentially available:
permitted high-excitation O\,{\sc i} lines,
the forbidden [O\,{\sc i}]
lines at 630\,nm, the OH vibrational-rotational and pure rotational
lines in the IR, and finally the OH electronic transitions in the UV
(which have not been included in the present analysis).
All have quite different sensitivities to the atmospheric structure
and the line broadening.
In addition, all have their pros and cons:
the [O\,{\sc i}] lines are very weak and likely blended
(Allende Prieto et al. 2001);
the O\,{\sc i} lines are sensitive to departures from LTE
%and thus the uncertain H-collisions 
(Kiselman 1993);
the OH IR lines are, like all molecular features,
very temperature sensitive (Asplund \& Garc\'{\i}a P{\'e}rez 2001);
the OH UV lines may be affected by missing opacity (Balachandran \& Bell 1998)
and possibly non-LTE effects (Asplund \& Garc\'{\i}a P{\'e}rez 2001).
Unfortunately the different lines imply wildly discortant abundances,
the difference amounting to about 0.3\,dex both for the
Holweger-M\"uller (1974) and {\sc marcs} (Asplund et al. 1997) 1D model atmospheres.
Normally the conclusions
from the OH IR have been embraced (log\,$\epsilon_{\rm O} \approx 8.90$),
largely due to the remarkably
small line-to-line scatter (Sauval et al. 1984).
An application of our new 3D model atmospheres to the problem, however,
suggests that the solar oxygen abundance may be significantly lower
(log\,$\epsilon_{\rm O} \approx 8.65$)
than previously thought (Allende Prieto et al. 2001;
Asplund et al., in preparation).

{\bf [O\,{\sc i}]:} Due to the exceptionally good agreement between
predicted and observed line profiles in general with 3D model atmospheres
(Fig. 3), disturbing blends are readily identified.
This statement naturally presumes that detailed comparisons have already been
performed for a range of lines with similar properties and typical
line formation regions, which fortunately is the case for [O\,{\sc i}].
The very poor agreement between the theoretical
and the observed profile, immediately signals that the [O\,{\sc i}] 630.0\,nm line
is indeed blended, which have occasionally been suspected but mostly
thought to be relatively unimportant. The effect of the blending Ni\,{\sc i} line
reduces the derived abundance significantly ($-0.13$\,dex), while the
transition from 1D to 3D model atmospheres and a revised $gf$-value are
of less importance ($-0.08$ and $-0.03$\,dex, respectively) but compounds the
overall difference. The forbidden line implies
a solar oxygen abundance of log\,$\epsilon_{\rm O} = 8.69 \pm 0.05$
(Allende Prieto et al. 2001).

{\bf O\,{\sc i}:} The O\,{\sc i} triplet around 777\,nm is well-known to
be susceptible to significant departures from LTE. We have therefore
investigated its line formation in detail with the 3D non-LTE code
{\sc multi3d} (Botnen \& Carlsson 1998; Asplund et al. 2002), employing
two independent snapshots from the 3D solar simulation and a
model atom consisting of 23 levels and 65 radiative transitions
(Kiselman 1993). The adopted model atom is sufficiently comprehensive for
this particular problem since even a two-level approach catches the
essentials of the line formation.
Collisional excitation and ionization due to electrons are accounted for
but not the corresponding case of hydrogen collisions, since
the available classical recipes (Drawin 1968) likely over-estimates
their importance by several orders of magnitude.
Our 3D non-LTE calculations
indeed reveal pronounced departures from LTE, in terms of abundance about
0.2\,dex. The dominant non-LTE mechanism is the same
as with 1D models, namely photon losses in the line itself.
%The non-vertical radiative transfer
%is in this case clearly of secondary importance.
Preliminary the triplet implies
a solar oxygen abundance of log\,$\epsilon_{\rm O} = 8.67 \pm 0.03$
(Asplund et al., in preparation).
%It should also be noted that the
%observed center-to-limb variation is well reproduced by our 3D calculations.

{\bf OH vibrational and rotational lines:} The slightly cooler temperature structure
in our 3D solar model compared with the standard Holweger-M\"uller (1974), which is
normally the 1D model of choice for solar abundance determinations, directly
translates to a lower derived oxygen abundance from the OH IR lines.
% given the great temperature sensitivity of molecule formation.
Equally important, however, is the presence
of temperature inhomogeneities due to the non-linear line formation. 
It therefore comes
as no surprise that 1D analyses typically will significantly 
over-estimate the abundances derived
from molecular lines. We have reanalysed 70 and 127 OH 
vibration-rotation and pure rotation
lines, respectively, which have previously been studied by 
Grevesse et al. (1984) and
Sauval et al. (1984) using up-to-date molecular line data.
The 3D line formation calculations are based on the assumption 
of LTE, which is likely
a reasonable approximation in view of the efficient collisional thermalization of
the relevant levels.
A potential source of error comes from the assumed chemical equilibrium in the OH
molecule formation, which would tend to artificially 
decrease the oxygen abundance
slightly. The magnitude of this effect is not known at this stage but will be
investigated further in the near future.
Keeping this in mind, preliminary unweighted means yield
log\,$\epsilon_{\rm O} = 8.61 \pm 0.03$ for the vibrational lines and
log\,$\epsilon_{\rm O} = 8.67 \pm 0.03$ for the rotational lines.

\begin{figure}
\plotfiddle{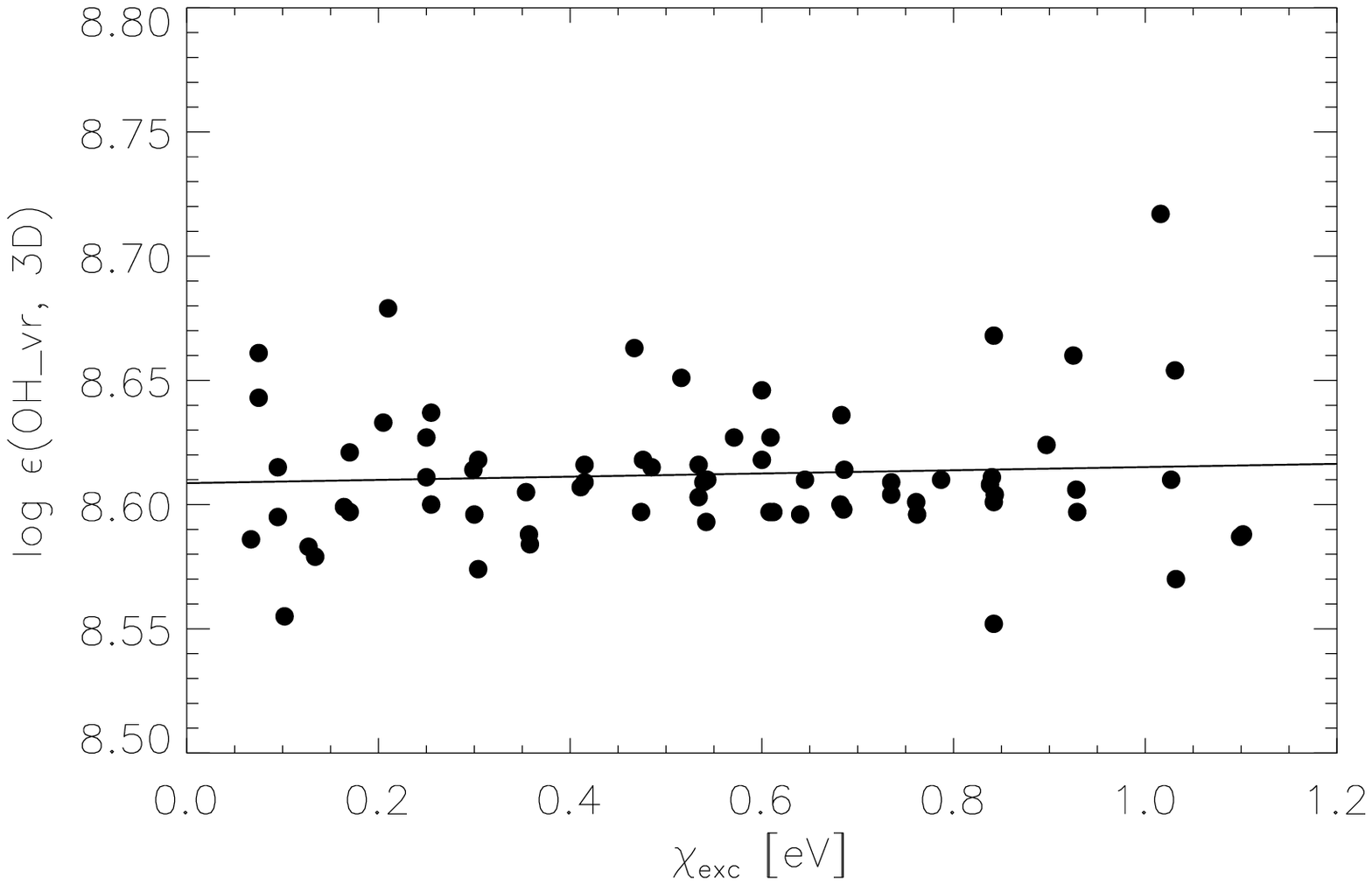}{8cm}{0}{75}{70}{-220}{-200}
\plotfiddle{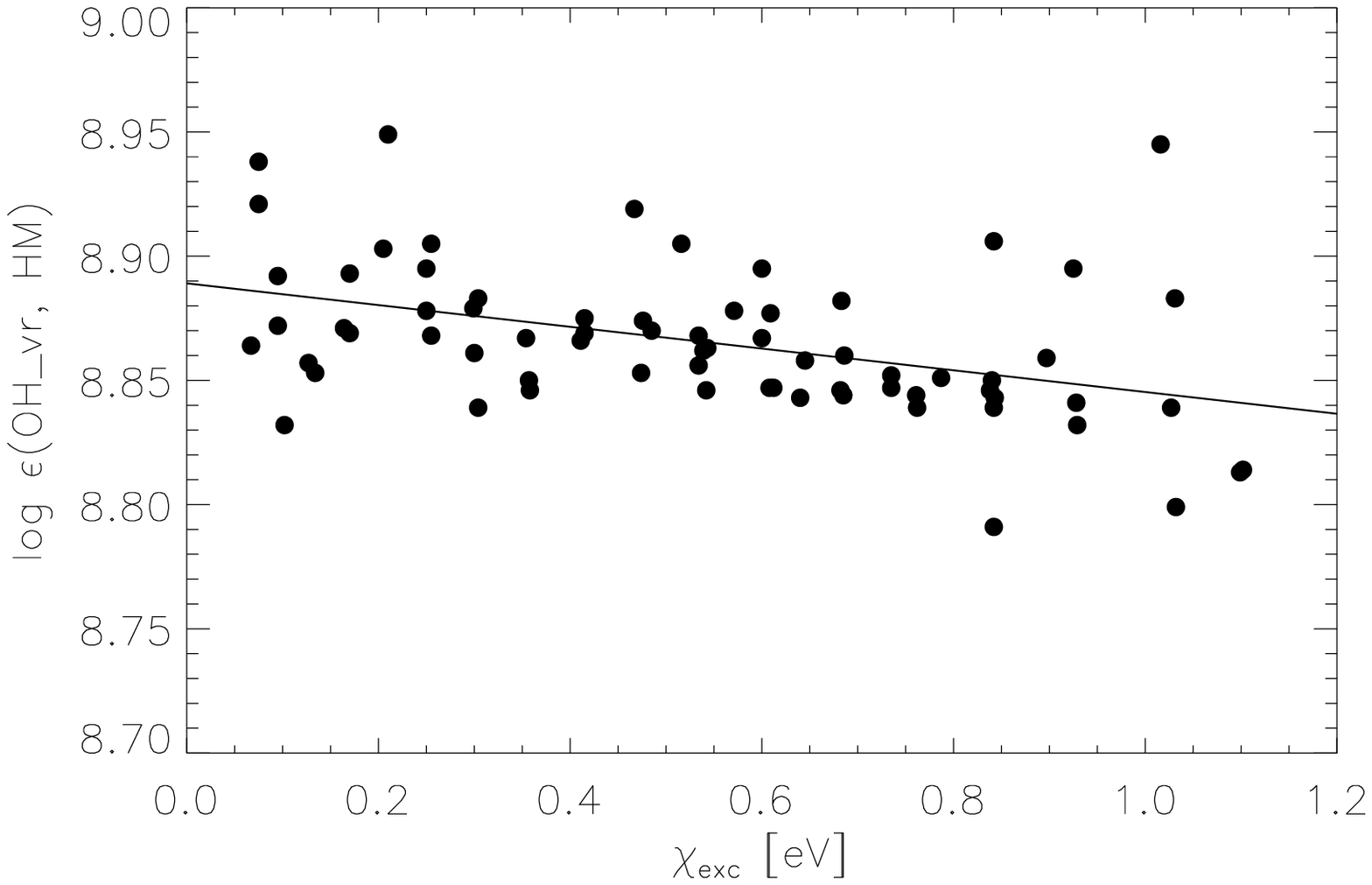}{8cm}{0}{75}{70}{-220}{-200}
%\plottwo{asplundm_f4a.ps}{asplundm_f4b.ps}
\caption{{\em Upper panel:} Derived O abundances from
the OH vibrational-rotational lines in the IR using a 3D hydrodynamical
simulation of the solar atmosphere. The solid line denotes the 
least-square-fit to the data.
{\em Lower panel:} Corresponding results for the 1D Holweger-M\"uller 
semi-empirical model atmosphere. Note the distinctly higher derived abundances
compared with the 3D case.
}
\end{figure}

{\bf Summary:} Although the exact value for the solar oxygen abundance depends
on the relative weights the different diagnostics are given, it is clear that
our detailed 3D investigation unanimously points to a significantly
lower solar oxygen abundance than previously thought based on classical
1D model atmospheres. Preliminary we estimate
log\,$\epsilon_{\rm O} = 8.65 \pm 0.05$  which is lower
by almost a factor of two.
Such a radical re-evaluation for the most abundant metal in the star
we know the best
is certainly a sobering realisation
when discussing derived stellar abundances based on inferior
spectra and where the modelling uncertainties can be much larger.
Our revision directly impacts also on
stellar evolution calculations and helioseismology among other topics.

\subsection{Oxygen in Metal-Poor Stars}

The oxygen abundances in stars born at different epochs
and locations carry vital information about the formation and
evolution of galaxies. In particular the temporal (with metallicity
as a practical although not perfect proxy) evolution of the oxygen-to-iron ratio
can yield estimates of the time-scales involved (O mainly from
SNe\,II while Fe largely from SNe\,Ia) as well as give important
clues to the nature of the elusive Pop III stars.
It has long been known 
%(Conti et al. 1967) 
that halo stars show an over-abundance of
oxygen ([O/Fe]\,$>0$) but an exact estimate has been
notoriously difficult to achieve due to the large discrepancies
between the different abundance indicators.
Typically the [O\,{\sc i}] line implies a more or less plateau
at [O/Fe]\,$\approx +0.4$ (Barbuy 1988; Nissen et al. 2002)
but due to weakness of the line
its use for [Fe/H]\,$<-2.0$ is restricted to analysis of
giants and subgiants.
Recent analyses of the OH UV lines in halo dwarfs on the other hand
seem to suggest an almost linear trend with metallicity, reaching
[O/Fe]$\approx +1.0$ at [Fe/H]$=-3.0$ (Israelian et al. 1998;
Boesgaard et al. 1999) while the OH IR vibrational lines in giants
again more support the conventional plateau (Melendez \& Barbuy 2002).
The confusing situation is completed by the fact that
the oxygen abundances derived from the
high-excitation O\,{\sc i} triplet at 777\,nm
tend to lay in between the OH UV and the [O\,{\sc i}] results.

We have recently applied the new generation of 3D hydrodynamical
model atmospheres of metal-poor stars (Asplund et al. 1999) in
an attempt to investigate whether a possible resolution to
the oxygen conundrum can be found.
Due to the lack of radiative heating
from spectral lines in halo stars, the adiabatic cooling from the rising gas in
the upflows dominate the energy balance. Consequently, the
temperature is much below the radiative equilibrium value
in the optically thin layers, which
is enforced in 1D models.
%This difference will naturally propagate over to the derived
%abundances from lines formed in these outer layers.

{\bf OH UV:} As noted before, molecule formation is extremely
temperature sensitive.
%, in the case of OH in typical line-forming
%conditions in metal-poor stars approximately $N_{\rm OH} \propto T^{-12}$
%(while still a trace species compared with O).
The lower temperatures
in 3D models therefore directly translates to much larger number densities
of molecules and thus to stronger lines.
The application of 3D model atmospheres appears to bring down the high
advocated linear trend to something which better resembles the
canonical plateau (Asplund \& Garc\'{\i}a P{\'e}rez 2001),
revealing a potential and very significant systematic error in
previous standard analyses using 1D model atmospheres.
This 3D LTE correction can easily amount to $0.6$\,dex or more
at [Fe/H]\,$=-3.0$.
%It should be noted, however, that the recent revision
%of the solar O abundance described above at least partly compensate
%the 3D corrections to published 1D investigations of OH UV lines.
The final word on this topic is not written yet however, as the
3D OH analyses sofar make the crucial assumption of LTE
both in the line formation and in the OH molecule formation.
Existing non-LTE calculations for molecules are very rare.
Asplund \& Garc\'{\i}a P{\'e}rez (2001) found the possibility of
significant departures from LTE in the OH line formation  but with little metallicity
dependence using a very
simplified two-level approach in 1D; the 3D case still remains to be investigated.
In addition, due to the rapid cooling of the upflowing material
it is not unlikely that the time available
is not sufficient for equilibrium chemistry to be established, in particular
for metal-poor stars with their very low temperatures in the upper layers.
%A very challenging project for the future is therefore time-dependent
%non-equilibrium chemistry coupled to non-LTE line formation.

{\bf [O\,{\sc i}]:} One could imagine that the [O\,{\sc i}] line should
be insensitive to the low surface temperatures in 3D models of halo stars
insofar it originates from the ground level of the ionization stage
where essentially all oxygen is residing. In reality though there are
significant downward 3D corrections also for this line, since
the line strength depends on the ratio of line and continuous opacity.
The latter is dominated by H$^{-}$ whose number density is proportional
to the free electrons which are much reduced in the low temperature gas
(Nissen et al. 2002). In terms of abundance the 3D corrections amount
to about $-0.2$\,dex at [Fe/H]\,$\approx -2.0$ for dwarfs and subgiants.
Whether a similar revision is needed for giants have to await the
construction of 3D models for such stars but slightly smaller corrections
would not be unexpected.

{\bf O\,{\sc i}:} The most problematic of the available diagnostics in
metal-poor stars is likely the O\,{\sc i} lines. Due to their high
excitation potential, the lines are formed in very deep layers where
the differences between 3D and 1D models are small. Already in 1D analyses,
the triplet yields higher O abundances than [O\,{\sc i}] which will be
aggrevated unless there are large metallicity-dependent
3D non-LTE effects on the triplet or
the adopted T$_{\rm eff}$-scale needs significant revision.
Preliminary 3D non-LTE calculations for the halo subgiant HD\,140283
(Asplund et al., in preparation) suggest
quite similar abundance corrections to the solar case, $\approx -0.2$\,dex,
which is apparently not enough (Nissen et al. 2002).
The best T$_{\rm eff}$-scale for halo stars likely comes from the IR flux method,
which gives too low T$_{\rm eff}$ to resolve the discrepancy even
with 3D models (Asplund \& Garc\'{\i}a P{\'e}rez 2001).
The use of 3D models may have resolved the problem between
the [O\,{\sc i}] and the OH UV lines but clearly not all answers are in
for the O\,{\sc i} triplet.

{\bf Summary:}
It seems like the adoption of the new generation of 3D hydrodynamical model
atmospheres holds a crucial key to the resolution of the long-standing problem
of oxygen abundances in halo stars. Although the final answer has not yet
emerged, the existing results together with the recent downward revision of
the solar O abundance described above point to a nearly flat [O/Fe] trend
but with a slow rise for [Fe/H]\,$\la -2$, i.e. somewhere in between the hotly contested
plateau and linear trends (Nissen et al. 2002).

\section{Concluding Remarks}

The above examples illustrate that existing 1D analyses may well be hampered
by significant systematic errors which can severely impact the inferred conclusions.
With the advent of a new generation of telescopes and
instrumentation it is clear that the dominant source of
error is no longer of observational nature but rather stems
from uncertainties inherent to the actual analyses.
There is obviously a great need for investing also significant
resources in improving the modelling aspects in order to place
the findings on firmer footing.
It is noteworthy that
the main obstacle for achieving this is simply a lack of manpower, in spite
of the existence of a very large community of astronomers relying on such
model atmospheres for their applications.

\acknowledgments

The work described in this contribution represent the
results of successful collaborations with a large number of
wonderful colleagues, in particular Carlos Allende Prieto, Mats Carlsson,
Remo Collet, Ana Garc\'{\i}a P{\'e}rez, Nicolas Grevesse, Dan Kiselman,
David Lambert, Poul Erik Nissen, \AA ke Nordlund,
Francesca Primas, Jacques Sauval, Bob Stein, and Regner Trampedach.
%Without their continuing efforts I would certainly not be in
%a position to present this report here.
Finally, I am grateful to the editors for their patience.


\begin{references}
\reference Allende Prieto, C., Lambert, D.L., Asplund, M.
2001, ApJ, 556, L65
\reference Allende Prieto, C., Asplund, M., Garc\'{\i}a L{\'o}pez, R.J.,
Lambert, D.L. 2002, ApJ, 567, 544
\reference Asplund, M. 2000, A\&A, 359, 755
\reference Asplund, M., Garc\'{\i}a P{\'e}rez, A.E. 2001, A\&A, 372, 601
\reference Asplund, M., Carlsson, M., Botnen, A.V. 2002, submitted to A\&A
\reference Asplund, M., Gustafsson, B., Kiselman, D., Eriksson, K. 1997,
A\&A, 318, 521
\reference  Asplund, M., Nordlund, \AA., Trampedach, R., Stein, R.F.
1999, A\&A, 346, L17
\reference  Asplund, M., Ludwig, H.-G., Nordlund, \AA., Stein, R.F.
2000a, A\&A, 359, 669
\reference  Asplund, M., Nordlund, \AA., Trampedach, R.,
Allende Prieto, C., Stein, R.F. 2000b, A\&A, 359, 729
\reference  Asplund, M., Nordlund, \AA., Trampedach, R., Stein, R.F.
2000c, A\&A, 359, 743
\reference Balachandran, S.C., Bell, R.A. 1998, Nature, 392, 23
\reference Barbuy, B. 1988, A\&A, 191, 121
%\reference Belyayev, A., Grosser, J.J.H., Menzel, T. 1999, Phys. Rev. A, 60, 2150
\reference Boesgaard, A.M., King, J.R., Deliyannis, C.P., Vogt, S. 1999 AJ, 117, 492
\reference Botnen, A.V., \& Carlsson, M. 1999, in
Numerical astrophysics, 
%ed. S.M. Miyama et al., 
379
%\reference Conti, P.S., Greenstein, J.L., Spinrad, H., Wallerstein, G.,
%Vardya, M.S. 1967, ApJ, 148, 105
\reference Dravins, D., Nordlund, \AA . 1990, A\&A, 228, 203
\reference Drawin, H.W. 1968, Z. Physik, 211, 404
%\reference Fleck, I., Grosser, J., Schnecke, A., Steen, W., Voigt, H. 1991,
%JPhysB, 24, 4017
\reference Grevesse, N., Sauval, A.J., van Dishoek, E.F. 1984, A\&A, 141, 10
\reference Gustafsson, B., Bell, R.A., Eriksson, K., Nordlund, \AA . 1975,
A\&A, 42, 407
\reference Holweger, H., M\"uller, E.A. 1974, Solar Physics, 39, 19
\reference Israelian, G., Garc\'{\i}a L{\'o}pez, R.J., Rebolo, R.
1998, ApJ, 507, 805
\reference Kiselman, D. 1993, A\&A, 275, 269
\reference Kiselman, D. 1997, ApJ, 489, L107
%\reference Kurucz, R.L. 1993, CD-ROM, private communication
\reference Mel{\'e}ndez, J., Barbuy, B. 2002, ApJ, 575, 474
\reference Mihalas, D., D\"appen, W., Hummer, D.G. 1988, ApJ, 331, 815
%\reference Nissen, P.E., Asplund, M., Hill, V., D'Odorico, S. 2000, A\&A, 357, L49
\reference Nissen, P.E., Primas, F., Asplund, M., Lambert, D.L. 2002, A\&A, 390, 235
\reference Nordlund, \AA. 1982, A\&A, 107, 1
\reference Nordlund, \AA., Dravins, D. 1990, A\&A, 228, 155
%\reference Nordlund, \AA., Spruit, H.C., Ludwig, H.-G., Trampedach, R. 1997,
%A\&A, 328, 229
\reference Sauval, A.J., Grevesse, N., Brault, J.W., Stokes, G.M., Zander, R. 1984, ApJ, 282, 330
\reference Stein, R.F., Nordlund, \AA. 1998, ApJ, 499, 914
\reference Uitenbroek, H. 1998, ApJ, 498, 427
\end{references}
\end{document}